\documentclass[aip,reprint, twocolumn]{revtex4-1}
\usepackage{amsmath}
\usepackage{graphicx}% Include figure files

\draft % marks overfull lines with a black rule on the right
\usepackage{xcolor}
\usepackage[english]{babel}

\begin{document}

%\preprint{}

\title{Dynamical steady-states of  active colloids interacting via chemical fields } %Title of paper



\author{Federico Fadda}
\email[]{f.fadda@uva.nl}
%\thanks{}
\affiliation{Institute of Physics, University of Amsterdam, 1098 XH Amsterdam, The Netherlands}

\author{Daniel A. Matoz-Fernandez}
\email[]{daniel.matoz@fuw.edu.pl}
%\thanks{}
\affiliation{Institute of Theoretical Physics, Faculty of Physics, University of Warsaw, Pasteura 5,
02-093 Warsaw, Poland}

\author{Ren\'e van Roij}
\email[]{r.vanroij@uu.nl}
%\thanks{}
\affiliation{Institute for Theoretical Physics, Center for Extreme Matter and Emergent Phenomena, Utrecht University, Princetonplein 5, Utrecht 3584 CC, The Netherlands}

\author{Sara Jabbari-Farouji}
\email[]{s.jabbarifarouji@uva.nl}
%\thanks{}
\affiliation{Institute of Physics, University of Amsterdam, 1098 XH Amsterdam, The Netherlands}


\date{\today}

\begin{abstract}
  We study the dynamical  steady-states of a monolayer of  chemically active self-phoretic colloids  as a function of packing fraction and self-propulsion speed by means of Brownian dynamics simulations. We focus on the case that a chemical field induces competing attractive positional and repulsive orientational interactions.  Analyzing the distribution of cluster size and local density as well as the hexatic order parameter, we distinguish four distinct dynamical states which include collapsed, active gas, dynamical clustering, and motility-induced phase-separated states. The long-range chemical field-induced interactions shift  the onset of motility-induced phase separation (MIPS) to very low packing fractions at intermediate self-propulsion speeds.  We also  find that the fraction of particles in the largest clusters is a suitable order parameter characterizing the  dynamical phase transitions from an active gas or dynamical clustering   steady-state to a phase-separated state upon increase of the packing fraction.  The order parameter changes discontinuously  when going from an active gas to a MIPS-like state at intermediate self-propulsion speeds, whereas it changes continuously at larger activities where the system undergoes a  transition from a  dynamical clustering state to MIPS-like state.   
  
\end{abstract}
 

\maketitle %\maketitle must follow title, authors, abstract and \pacs

\section{Introduction}

Active matter systems exhibit characteristics that are radically distinct from the ordinary passive materials that we deal with in our daily life. The term active matter refers to  collectives of living and nonliving units  that consume energy to generate a type of autonomous motion maintaining the system constantly far from equilibrium.~\cite{Ramaswamy2010, marchetti1,stark5} Interestingly, when the active units interact, spectacular collective phenomena can emerge that have no counterparts in equilibrium systems. Striking examples include the emergence of collective motion,~\cite{Vicsek,Chate2020,Liebchen2017} giant number fluctuations~\cite{narayan2007long, Fily2012} and formation of clusters of particles interacting with a gaseous background, also known as motility-induced phase separation.~\cite{MIPS-1,MIPS-3,lowen5,Damme2019} %\cite{MIPS-1,MIPS-2,MIPS-3,MIPS-4,MIPS-5,lowen2,lowen3,lowen4,lowen5}. %However, the types of behaviors and the factors that control them remain elusive for most systems .
%At the microscopic scale, the active system's force-generating elements are supplied with energy from their surroundings.
Continuous dissipation of energy  at the level of individual units supplied by the  surroundings  leads to the spontaneous emergence of large-scale collective dynamics and complex hierarchical structures like those found in living systems. The energy flow within active systems leads to novel symmetries, conservation laws, and material properties completely different from those found in equilibrium systems.~\cite{ Ginot2015, solon2015pressure, theurkauff2, banerjee2017odd, scheibner2020odd,PhysRevX_22}

%For example, active systems may arrange into a macroscopically migrating flock, produce spontaneous steady-state oscillations and segregate in the presence of purely repulsive interactions in what has become known as motility-induced phase separation

Active agents not only use the environment as a fuel but also for navigation. Living organisms, for example, continuously exploit this mechanism to sense specific chemical substances and respond accordingly. For instance, chemoattraction processes can help microorganisms to find food and nutrients, whereas chemorepulsion directs an organism away from  harmful chemicals such as poisons and toxins \cite{stark4,liebchen1,lowen7,golestanian10} or predators. \cite{lowen6,liebchen2} In the motion of bacterium \textit{E. Coli} chemotaxis is visible in the \textit{run and tumble} motion of a bacterium which alternates between self-propulsion in a fixed direction (run) and  reorientation of its direction (tumble) in search of nutrients.~\cite{valeriani} 
Experiments have also revealed that chemotaxis allows self-propelled droplets and cells to efficiently move in a patterned environment like a maze.~\cite{maass2,cell} Moreover, chemotaxis is also a key element for cell mechanics, enzymes, and elastic shells. ~\cite{golestanian11,golestanian9,daniel1}\\
\hspace{1cm}The collective organization of self-driven organisms across all scales has inspired soft-matter scientists to build artificial self-propelled systems with controlled properties to understand the basic features of structure formation in nonequilibrium chemically active systems.  A prototypical example is  self-propelled Janus colloidal systems driven by self-diffusiophoresis, which have been designed to mimic  microorganisms like bacteria. They create a \emph{nonuniform chemical field} around themselves and initiate diffusiophoresis which mediate long-range  chemo-phoretic interactions among them.\\
\hspace{1cm}From a theoretical point of view, various theoretical models have been proposed for chemotaxis through the years. From continuum models such as well-established phase field \textit{Keller-Segel} model \cite{keller_segel2,liebchen1} to particle-based simulations which couple the dynamics of particles to the dynamics of relevant chemical fields.~\cite{stark1,stark2,liebchen3,golestanian,liebchen6} 
 In the simplest case, the Active Brownian Particles (ABP) model \cite{Fily2012, lowen2020inertial} is generalized to couple the particles  orientational degrees  of  freedom  to a chemical field~\cite{liebchen3}. 

The particle-resolved studies so far have focused on low packing fractions $\Phi <0.2$.~\cite{stark1,stark2,liebchen3} To our knowledge, the interplay between crowding effects and phoretic interactions has not been explored so far.
In this paper, we offer  one of the first investigations into how density affects the dynamical   steady-states of chemotactic active colloids. We have focused on the chemotactic active colloid model~\cite{stark1,stark2} developed by by Pohl and Stark.  
%~[\onlinecite{stark1}] and [\onlinecite{stark2}]. 
To overcome the challenge of simulating  a large number of particles, we have introduced the chemical-mediated interactions in an efficient Brownian dynamics GPU implemented code~[\onlinecite{MatozKITP}], which allows for large-scale simulations. Employing this code, we have investigated the structural and dynamical features of two-dimensional self-phoretic colloids for the case that that chemical field induces competing attractive positional and repulsive orientational interactions. We have investigated the dynamical  steady-states of the system  upon varying the self-propulsion speed and packing fraction.    

The remainder of the article is organized as follows. In Section \ref{sec:simulations}, we describe the physical model for chemotactic particles and  numerical details of our particle-based simulations. In Section \ref{sec:steady_states}, we  present our state diagram as a function of reduced self-propulsion speed and  packing fraction. Then, we  discuss the  distinct signatures of each dynamical  steady-state.  In section  IV, we investigate the characteristics of dynamical phase transitions by analyzing various order parameters including the fraction of particles in the largest cluster and hexatic order parameter. 
Finally, we conclude our work in Section V  with a summary of our most important finding and directions for future work.

%an  asymmetric distribution of chemical products in Janus colloids drives a phoretic slip velocity in the interfacial layer of colloid which leads to their self-propulsion velocity with speed $v_0$ along direction $\textbf{e}$, which we assume  to be fixed in the particle frame. 

\section{Simulation details} \label{sec:simulations}

\subsection{Dynamical equations}

\begin{figure}
	%\centerline{\includegraphics[width=0.85\textwidth]{figures/cartoon.png}}
\centerline{\includegraphics[width=0.95\linewidth]{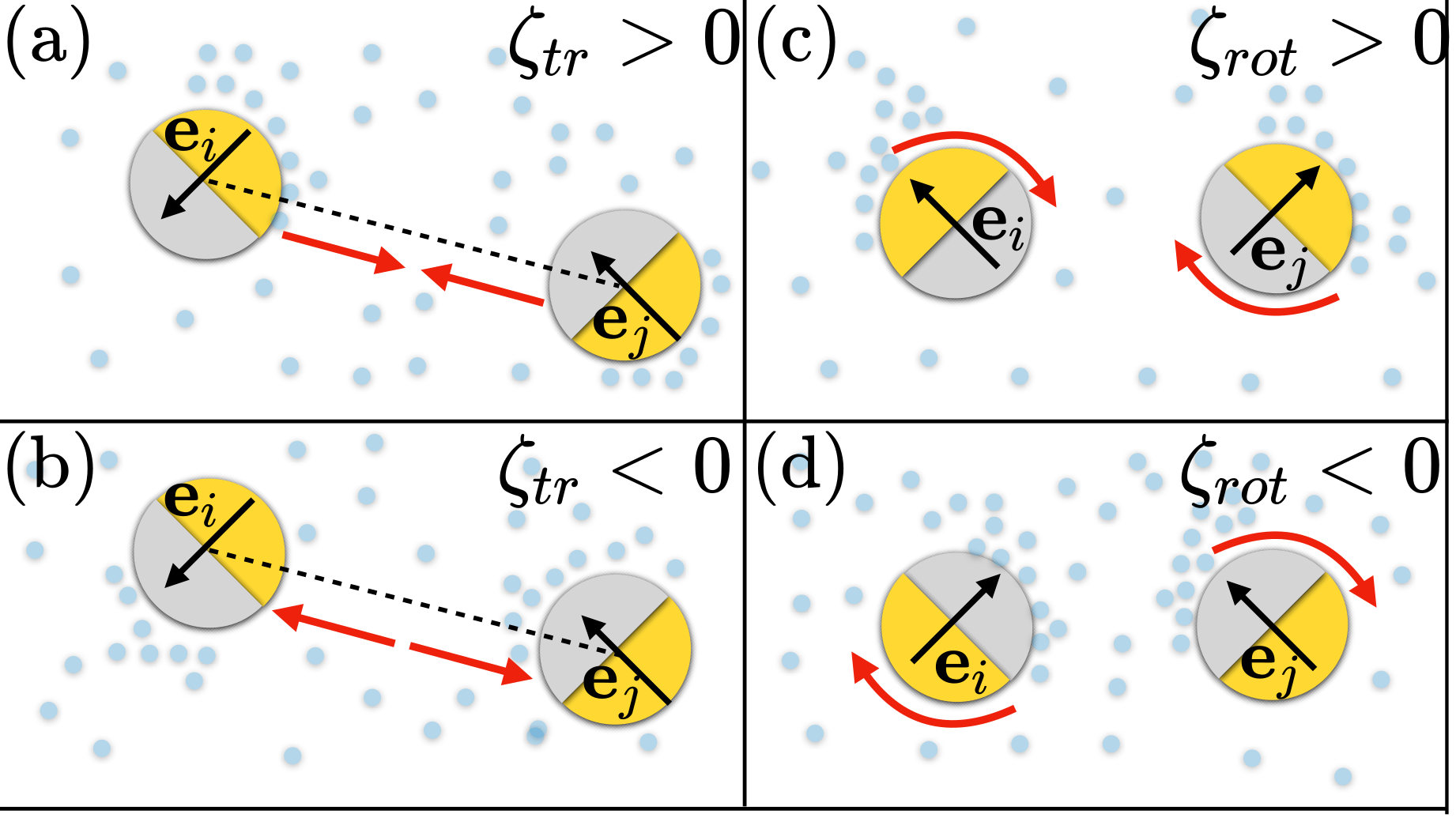}}
\caption{Schematics of  chemical field mediated  interactions between self-catalytic Janus colloids for different cases  of translational $\zeta_{\text{tr}}$ and rotational $\zeta_{\text{rot}}$ chemotactic mobility  parameters in the situation that the colloids act as chemical sinks. 
%The diffuse blue particles represent  the chemicals. 
(a)  When $\zeta_{\text{tr}}>0$, the colloids move towards each other, (b) when $\zeta_{\text{tr}}<0$, the colloids move away from each other, (c) when $\zeta_{\text{rot}}>0$ the colloids rotate towards each other, and (d) when $\zeta_{\text{rot}}<0$ they rotate way from each other. }
\label{fig0}
\end{figure}

Following earlier work on active colloids interacting via diffusiophoretic interactions,~\cite{stark1,stark2} we focus on a system of $N$ self-phoretic spherical colloids of radius $a$ confined in a two-dimensional square box. In terms of experiments, this model represents heavy catalytic Janus colloids that settle to the bottom of the experimental cell to create a colloidal monolayer. The chemically-active parts of the Janus colloids  initiate reactions which generate chemical products.
Due to the asymmetric distribution of the chemical products around  a catalytic Janus colloid labelled by index $i$, it self-propels at a speed $v_0$ along direction $\mathbf{e}_i$ fixed in the particle frame (Fig. \ref{fig0}), leading to a phoretic slip velocity at the particle's interfacial layer.  In principle, $v_0$ depends on the chemical concentration, however, when chemicals fuels are abundant, one expects $v_0$ not to change noticeably due to local inhomogeneities of chemical field $c(\mathbf{r})$ and, thus, we consider it as a fixed control parameter.~\cite{theurkauff,stark1,stark2} 
In addition, the same chemical gradient field  $ \boldsymbol{\nabla} c(\textbf{r})$ that provides the drive for the propulsion of colloids, induces translational and rotational drift velocities which mediate diffusiophoretic interactions between chemical-consuming colloids.
Under these considerations, for the case of half-coated Janus colloids, the translational and rotational drift velocities then can be written as~\cite{anderson}
\begin{equation}
\begin{split}
\textbf{v}_{D,i}&=-\zeta_{\text{tr}}\boldsymbol{\nabla} c(\textbf{r}_i),\\
 {\mathbf \omega}_{D,i} & =-\zeta_{\text{rot}} \textbf{e}_i \times \boldsymbol{\nabla} c(\textbf{r}_i), 
\end{split}
\end{equation}
where $\zeta_{\text{tr}}$ and $\zeta_{\text{rot}}$ are the translational and rotational phoretic mobility coefficients that encode the active colloids' response to the chemical field gradients. If $\zeta_{\text{tr}}>0$ colloids move away from the direction of local chemical gradient while for $\zeta_{\text{tr}}<0$ the particles move towards regions of higher concentration of chemicals. Likewise, for $\zeta_{\text{rot}}>0$, the colloids rotate away from the direction of the local gradient, whereas for $\zeta_{\text{rot}}<0$ they reorient in the direction of the chemical gradient (See Fig. \ref{fig0}).
%the $i$-th particle is directed towards the neighbours corresponding to effective attraction, while $\zeta_{\text{tr}}<0$ is an effective repulsion. In a similar way, $\zeta_{\text{rot}}>0$ rotates the direction of the ith particle towards the neighbours. Therefore also $\zeta_{\text{rot}}>0$ ($\zeta_{\text{rot}}<0$) acts as effective attractive (repulsive) term.

%\cite{saintillan1,lushi,nejad,zottl,kapral}
In what follows, we neglect the hydrodynamic interactions and model the collective motion of phoretically interacting active colloids with positions $\textbf{r}_{i}=x_{i} \hat{\mathbf{e}}_x + y_{i} \hat{\mathbf{e}}_y$ and orientations $\textbf{e}_{i}=\cos \varphi_{i} \hat{\mathbf{e}}_x + \sin \varphi_{i} \hat{\mathbf{e}}_y$ in the overdamped limit, describing it by the following Brownian dynamics equations ~\cite{stark1,stark2,stark3,stark6,liebchen7}
\begin{equation}
	\dot{\textbf{r}}_{i}=v_{0}\textbf{e}_{i}+\textbf{F}_{i}/\gamma_{\text{tr}}-\zeta_{\text{tr}}\boldsymbol{\nabla} c (\textbf{r}_{i}) +\sqrt{2D_{\text{tr}}}\,\boldsymbol{\xi}_{\text{tr},i},
	\label{pos}
\end{equation}
\begin{equation}
	\dot{\textbf{e}}_{i}=-\zeta_{\text{rot}}(\textbf{1}-\textbf{e}_{i}\boldsymbol{\otimes} \textbf{e}_{i})\boldsymbol{\nabla} c(\textbf{r}_{i})+\sqrt{2D_{\text{rot}}}\, \boldsymbol{\xi}_{\text{rot},i} \times \textbf{e}_{i},
	\label{dir}
\end{equation}
where $\gamma_{\text{tr}}$ is the translational drag coefficient, and 
$\boldsymbol{\xi}_{\text{tr},i}$ and $\boldsymbol{\xi}_{\text{rot},i}$ are translational and rotational white noises with zero mean and unit variance, \emph{viz.}, $\langle \boldsymbol{\xi}_{\text{tr},i}(t)\boldsymbol{\otimes}\boldsymbol{\xi}_{\text{tr},i}(t')\rangle= \textbf{1}\delta(t-t')$ and $\langle \boldsymbol{\xi}_{\text{rot},i}(t)\boldsymbol{\otimes} {\boldsymbol{\xi}_{\text{rot},i}}(t')\rangle= \textbf{1}\delta(t-t')$. 
The term $\textbf{F}_{i}=-\nabla\sum_{j\neq i}U^{\text{WCA}}(r_{ij})$ corresponds to the force on the $i$-th particle due to excluded volume repulsive interactions from all the other  particles which is  modelled by the Weeks-Chandler-Anderson potential,~\cite{WCA}
\begin{equation}
 U^{\text{WCA}}(r)=4\epsilon\left [ \left ( \frac{\sigma}{r} \right )^{12} - \left ( \frac{\sigma}{r} \right )^{6}\right ]+\epsilon \quad r<2^{1/6}\sigma,   
\end{equation}
with $\sigma=2a$ the particle diameter, and $\epsilon$ sets the strength of the potential. Note that, unlike the Refs. \cite{stark1,stark2,stark3} where the authors implemented a manual repulsion between particles separating them along the line connecting their centers in case of overlap $r<2a$, we introduce explicitly the WCA potential to account for the particles excluded volume.
As can be seen from Eqs.~\eqref{pos} and \eqref{dir}, the self-propelled particles are coupled to the chemical field $c(\textbf{r},t)$ that represents the coarse-grained continuum concentration of the involved chemical species at time $t$.
We assume that the chemical field diffuses in the infinite three-dimensional half-space with a diffusion coefficient $D_{\text{c}}$ and has sinks at the positions of the particles since they consume the chemical at rate $k$ such that $c(\textbf{r},t)$  satisfies the following reaction diffusion equation
\begin{equation}
	\dot{c}(\textbf{r})=D_{\text{c}}\nabla^{2}c-k\sum_{i=1}^{N}\delta(\textbf{r}-\textbf{r}_{i}).
\end{equation}
Since the active colloids consume the chemicals, for the case
$\zeta_{\text{rot}}>0$ particles rotate away from chemical sinks (other colloids), thus giving rise to an effective interparticle repulsion whereas  $\zeta_{\text{rot}}<0$ leads to an effective, attractive alignment between colloids.
Likewise, $\zeta_{\text{tr}}>0$ ($\zeta_{\text{tr}}<0$) leads to an effective attraction between colloids as the particles move towards (away from) the neighbor sink while moving away from (towards) the concentrated regions of chemicals. %corresponding to effective attraction, while $\zeta_{\text{tr}}<0$ is an effective repulsion. In a similar way, $\zeta_{\text{rot}}>0$ rotates the direction of the ith particle towards the neighbours. Therefore also $\zeta_{\text{rot}}>0$ ($\zeta_{\text{rot}}<0$) acts as effective attractive (repulsive) term.
Typically, the chemical field diffuses much faster than the colloids such that $D_c \gg D_{\text{tr}}$. Hence, we can neglect the time dependence of the chemical field equation and adopt a stationary solution given by the Poisson equation and its solution in 3D,
%The stationary equation is identical to the Poisson equation, and its solution in 3D is given by
\begin{equation}
	c_{3d}(\textbf{r})=c_{0}-\frac{k}{4 \pi D_{\text{c}}}\sum_{i=1}^{N}\frac{1}{|\textbf{r}-\textbf{r}_{i}|},
	\label{poisson}
\end{equation}
where $c_{0}$ is the background chemical concentration. Note that Eq.(\ref{poisson}) implies that each colloid instantly establishes a stationary long-range chemical sink around itself, which moves with it. The effective two-dimensional concentration field, in which the colloidal monolayer lives, can be approximately obtained by integrating over a thin layer of thickness $h \approx 2a$  yielding $c_{2d}(\textbf{r})=h c_{3d}(\textbf{r})$.
%Strictly speaking, we would need to implement a non-flux boundary condition at the boundary of the infinite half-space, but this will not change the basic 1/r dependence of the chemical field. To introduce a two-dimensional concentration field, in which the colloidal monolayer moves, we integrate over a thin layer of thickness h = 2a and obtain c2d(r) =   h 0 c3d(r)dz ≈ hc3d.
%Finally, the chemical concentration field reads

\subsection{Dimensionless equations of motion}
To carry out many-body simulations, we first render the equations  Eqs.~\eqref{pos}, \eqref{dir} and  \eqref{poisson}  dimensionless.
Similarly to references,~\cite{stark1,stark2,stark3} we choose  $t_{r}=1/(2D_{\text{rot}})$ and $l_{r}=\sqrt{D_{\text{tr}}/D_{\text{rot}}}$ as units of time and  length, respectively, defining dimensionless length and time units as $\textbf{r}^*=\textbf{r}/l_r$ and $t^*=t/t_r$. We also choose $\epsilon$ as the unit of energy. The equations of motion in reduced units  become
\begin{equation}
	\dot{\textbf{r}}_{i}^*=Pe \, \textbf{e}_{i}+\textbf{F}_{i}^*/\gamma_{\text{tr}}^*-\zeta_{\text{tr}}^*\boldsymbol{\nabla} c^* (\textbf{r}_{i}^*) +\boldsymbol{\xi}_{\text{tr},i}^*,
	\label{eq:tr}
\end{equation}
\begin{equation}
	\dot{\textbf{e}}_{i}=-\zeta_{\text{rot}}^*(\textbf{1}-\textbf{e}_{i}\boldsymbol{\otimes} \textbf{e}_{i})\boldsymbol{\nabla} c^*(\textbf{r}_{i}^*)+ \boldsymbol{\xi}_{\text{rot},i}^* \times \textbf{e}_{i},
	\label{eq:rot}
\end{equation}
in which $\textbf{F}^*_i=\textbf{F}_i \sqrt{D_{\text{tr}}/D_{\text{rot}}}/\epsilon$, $\gamma_{\text{tr}}^*=2 \gamma_{\text{tr}} D_{\text{tr}} /\epsilon $, $\boldsymbol{\xi}_{\text{tr,rot},i}^*=\boldsymbol{\xi}_{\text{tr,rot},i}D_{\text{rot}}^{-1/2}$. The reduced two-dimensional concentration field is defined as $c^*=4 \pi c D_{\text{c}}\sqrt{D_{\text{rot}} } /(kh\sqrt{D_{\text{tr}}})$. The essential dimensionless parameters appearing in the reduced equations include
the reduced self-propulsion speed called P\'{e}clet number $Pe=v_{0}/(2\sqrt{D_{\text{tr}}D_{\text{rot}}})$, $\zeta_{\text{tr}}^*=\zeta_{\text{tr}}k h D_{\text{rot}}^{1/2}/(8\pi D_{\text{c}}D_{\text{tr}}^{3/2})$ and $\zeta_{\text{rot}}^*= \zeta_{\text{rot}} k h/(8\pi D_{\text{c}}D_{\text{tr}})$.

To summarize,   four dimensionless  parameters determine the collective dynamics of self-phoretic colloids: the P\'{e}clet number $Pe$, the two chemotactic constants $\zeta_{\text{tr}}^*$ and $\zeta_{\text{rot}}^*$, and the packing fraction $\Phi=N\pi a^{2}/L^{2}$ with $L$ being the linear size of the system. 
We note that the definition of $Pe$ used here is different from the typical studies of Active Brownian Particles (ABP) where it is defined as $Pe^{\text{ABP}}=v_{0}\sigma/D_{\text{tr}}$.~\cite{hagan,theurkauff}  Usually, it is assumed that the two diffusive coefficients follow the equilibrium  relation $D_{\text{rot}}= 3D_{\text{tr}}\sigma^{2}$.  Employing this relation, we find that  $Pe^{\text{ABP}}=2 \sqrt{3} Pe$. Thus the two P\'{e}clet number definitions only differ by a numerical prefactor and can be easily mapped to each other for the sake of comparison with the literature.

\subsection{Implementation of Brownian dynamics simulations}

To integrate the many-body equations of motion Eqs.~\eqref{eq:tr} and \eqref{eq:rot} we employed the Euler-Maruyama scheme.~\cite{leimkuhler2015} Therefore, the positions and angles are evolved during a time step $dt$ according to:
%\iffalse
\begin{equation}
\textbf{r}_{i}(t+dt)=\mathbf{r}_{i}(t)+dt Pe\, \textbf{e}_{i}-\zeta_{\text{tr}}\boldsymbol{\nabla} c(\textbf{r}_{i})+dt\mathbf{F}_{i}/\gamma_{\text{tr}}+\sqrt{dt}\boldsymbol{\mathcal{N}}_{\text{tr},i},
\label{pos1}
\end{equation}
\begin{equation}
\varphi _{i}(t+dt)=\varphi_{i}(t)-dt\zeta_{\text{rot}}[-\sin\varphi_{i},\cos\varphi_{i}]\cdot\boldsymbol{\nabla} c(\mathbf{r}_{i})+\sqrt{dt}\mathcal{N}_{\text{rot},i},
\label{dir1}
\end{equation}
where for the ease of notation we have dropped the $^*$ superscripts from the dimensionless quantities and we will continue to do so in  what follows.
Here $\mathcal{N}_{\text{tr},i}$ and $\mathcal{N}_{\text{rot},i}$ denote the random variables  representing the Gaussian white noise. They are generated for each colloid at each time step using a normal distribution with a mean of zero and a standard deviation of one, respectively.
%\fi
To increase the efficiency of large-scale simulations, we added the chemical fields to a CUDA code for particle-based models.~\cite{MatozKITP}
%One demanding part of these simulations is the computation of all collisions among the particles. To solve this, we implement a neighbour-list method to identify only the close particles with a mutual distance below $1.12\sigma$. 
One demanding part of the simulations concerns the evaluation of the long-range chemical interactions between particles, where one needs to calculate the gradient of the chemical as the sum of $N-1$ terms of the from $1/|\mathbf{r}_{i}-\mathbf{r}_{j}|$ for each particle $i$. To solve for this, we have used a fast $N$-body algorithm that makes use of shared memory.~\cite{gpugems3} In addition, to reduce further the computational costs, we updated the chemical field every $50$ time-step.
%Contrarily to the Lennard-Jones interaction, chemical sensing is not only limited to the close particles but to any of them. 
Within clusters of active colloids, the concentration field
cannot freely diffuse. This leads to screening of the chemical field which is not taken into account by the equations of motion.
To account for it, we follow the Stark's group recipe  \cite{stark1,stark2,stark3} and introduce a manual screening rule whenever a colloid is surrounded by six closely packed neighbors,{\it i.e.}, when colloid has six neighbors all located at a distance $r \le \xi$ from the particle, with $\xi$ defining a screening length. In this situation, we replace  the term $1/r$ in Eq.~\eqref{poisson}  with $\exp[-(r-\xi)/ \xi]$, in which  $r=|\textbf{r}_j-\textbf{r}_{i}|$ and $\xi=2a(1+\delta)$ with $\delta=0.3$.\\

%\sara{{\it i.e.}, there exist six neighbors all located at a distance $r \le 1.22 \sigma$ from the particle.} In this situation, we replace  the term $1/r$ in Eq.~\eqref{poisson}  with $\exp[-(r-\xi)/ \xi]$, in which  $r=|\textbf{r}_j-\textbf{r}_{i}|$ and $\xi=2a(1+\delta)$ is the screening length with $\delta=0.3$.\\

%

\subsection{Simulation parameters}
In our simulations, we fix the chemotactic mobility parameters to $\zeta_{\text{tr}}=15.4$ and $\zeta_{\text{rot}}=-0.38$ while varying the P\'eclet number $Pe$ and packing fraction $\Phi$ in the ranges $5 \le Pe \le 30$ and $0.001 \le \Phi \le 0.7$.    
 We chose to set $N=10^{4}$ which allows for an efficient scan of the phase diagram in a reasonable time using a time-step in the range $dt=10^{-5}- 5 \times 10^{-5}$. We observed that the  steady-state is reached after $t=(1-5)\times10^4$ in scaled units depending on the packing fraction. For small $\Phi$ and $Pe$ the  steady-state time is naturally increased. As in Ref. ~[\onlinecite{stark1, stark2, stark3, stark4}], the system is enclosed by impenetrable walls. Therefore, when particles collide with the walls (i.e. $\vert \mathbf{r}_i - \mathbf{r}_{wall}\vert<a$), they are reflected randomly away from them into the simulation box.
 
 %\sara{when particles collide with the walls, they are reflected randomly away from them. Daniel, could you please clarify how you define collision with the wall?}
 
  %We initialized the system with $N=10^{4}$ particles arranged in a square lattice and then the system was left to evolve reaching a final  steady-state. 
 %The system is enclosed by walls with particles changing randomly their directions inside the box once they hit them.\\
 
 \begin{figure*}
	%\centerline{\includegraphics[width=0.85\textwidth]{figures/phase-diagram.pdf}}
\centerline{\includegraphics[width=0.95\textwidth]{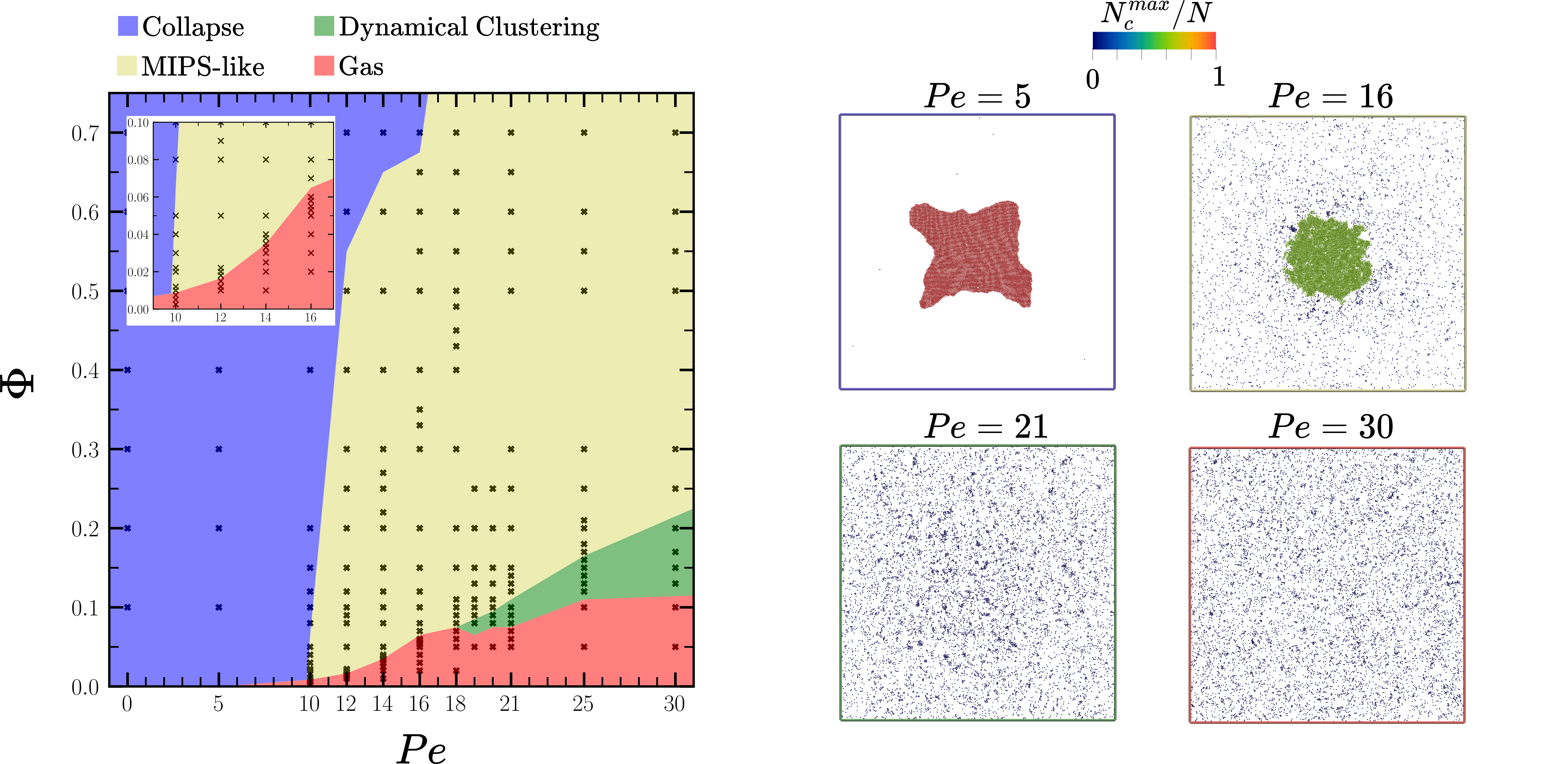}}
	
	\caption{State diagram of self-phoretic colloids in the P\'eclet-packing fraction  $(\Phi$-$Pe)$ representation for translational and rotational mobility coefficients $\zeta_{\text{tr}}=15.4$ and $\zeta_{\text{rot}}=-0.38$, respectively.  We distinguish four distinct dynamical states: active gas (red region), dynamical clusters (green region), MIPS-like (yellow region) and collapsed state (blue region). The right panel shows representative snapshots for each dynamical state but different, in which the largest cluster is color coded by $ N_{\text{c}}^{\text{max}}/N$.  The examples are chosen for a fixed $\Phi=0.1$ and different $Pe$ values as given on the top of each snapshot.}
	
	\label{phase-diagram_2}
\end{figure*}
 
 \subsection{Computed quantities}
\subsubsection*{Clustering algorithm}

To determine the clusters in the system, we first identify the neighbours of each particle within a threshold cutoff $r_c$ by using $k$-d trees.~\cite{cormen2009Algorithms} Then, we construct an undirected graph that labels the connected particles by cluster-$id$ and cluster-$size$.~\cite{SciPy, NetworkX} Using this information then is easy to construct several useful quantities. We first calculated the normalized  the distribution $P(n)$ of clusters of any size $n$ as $P(n)=N(n)/\sum_{n=1} N(n)$, where $N(n)$ is the number of clusters of  any size $n$.   
The mean cluster size for any snapshot is given by 
\begin{equation}
  N_{\text{c}} (t)= \sum_{n=1}n P(n) 
\end{equation}
where $\sum_{n=1} P(n)=1$. In addition, we obtain the time-averaged cluster size $ N_{\text{c}}^{\text{avg}}$, and the fraction of particles in largest cluster $N_{\text{c}}^{\text{max}}$ by  averaging over many steady-state snapshots. The normalized maximum cluster in the system $N_{\text{c}}^{\text{max}}$ to the total number of particles $N$ can be used as an order parameter~\cite{Chandra2012,MatozPerco,dipolar_active_abp_2d_klapp} characterizing the transition from gas or dynamical clustering regimes to the phase-separated state.

\subsubsection*{Computation of local density distribution} 
 
Following Ref.~[\onlinecite{MIPS-3}], we discretize the system by dividing it into squares of linear size $\xi$, so that continuous space is now replaced by a discrete lattice containing $L^2/\xi^2$ sites. Then, we construct a discrete density, $\bar{\phi}(\mathbf{r})$, defined for discrete positions $\mathbf{r}$ located at the center of each grid square given by
\begin{equation}
    \bar{\phi}(\mathbf{r}) = \frac{1}{2\pi\,\xi^2} \sum_{i=1}^N \Theta(\xi - \vert \mathbf{r}-\mathbf{r}_i\vert)
\end{equation}
 where $\Theta(y)$ is the Heaviside function, and $\xi=[2.0,4.0]\sigma$.
%\sara{Daniel, can you clarify what range of value of $\xi$ you used?}
 \subsubsection*{Hexatic order parameter}
The global hexatic order parameter, also known as six-fold bond orientational order parameter, is defined as\cite{stark2}
\begin{equation}
q_{6}=\left | \frac{1}{N}\sum _{k=1}^{N}q_{6}^{(k)} \right | %\in [0;1]
\end{equation}
with
\begin{equation}
q_{6}^{(k)}=\frac{1}{\mathcal{Z}_{6}^{(k)}}\sum _{j\in\mathcal{Z}_{6}^{(k)}}^{N}e^{i6\alpha_{kj}} .
\end{equation}
Here, $\mathcal{Z}_{6}^{(k)}$ are the number of nearest neighbours of particle $k$ and $\alpha_{kj}$ is the angle between the vector connecting particle $k$ to $j$ and the horizontal $x$-axis. Given this definition, the parameter $q_{6}^{(k)}=1$  if a particle is surrounded by $6$ closely packed neighbours in a system with perfect hexatic order.

\section{Dynamical steady-states of chemotactic colloids} \label{sec:steady_states}

\subsection{State diagram $Pe-\Phi$}
We start by giving an overview of our state diagram as a function of  the dimensionless self-propulsion speed, {\it i.e.}, the P\'{e}clet number ($Pe$) and packing fraction ($\Phi$) while keeping the  translational and rotational phoretic mobility parameters constant.
We fixed the values of translational and rotational phoretic mobility parameters to  $\zeta_{\text{tr}}=15.4$ and $\zeta_{\text{rot}}=-0.38$. For this choice of parameters, each particle  rotates away from  other chemicals consuming the colloids, whereas it translationally   moves towards other colloids. Overall, active colloids experience effective Coulomb-like phoretic attractive forces  $\propto -\zeta_{\text{tr}}/r^2$,~\cite{palacci, Palacci2014} and repulsive torques $\propto -\zeta_{\text{rot}}/r^2$.
Prior studies of low density colloids at $\Phi=0.05$ for this set of parameters show that  colloids form dynamical clusters.~\cite{stark1,stark2}
%In what is next, we will discuss the distinct characteristics of each motility pattern.

 Varying $Pe$ and $\Phi$, we observe four distinct dynamical \emph{ steady-states} as summarized in the state diagram of Fig.~\ref{phase-diagram_2}. The dynamical states include  an active gas, a  collapsed state,  a dynamical clustering state, and a phase separation into a dense large cluster coexisting with a dilute active gas.  Representative snapshots of the configuration of each dynamical state are presented in the right part of Fig.~\ref{phase-diagram_2}. In a collapsed state, all the  particles collapse into a giant cluster. Conversely, in a dynamic clustering state,  we observe clusters of finite size with mean cluster size  larger than three where particles actively join and leave the clusters. Finally, in the phase-separated state, a big cluster coexists with dynamical clusters similar 
 to the \emph{Motility-Induced Phase Separation} (MIPS) observed for purely repulsive active colloids~\cite{hagan} at sufficiently high self-propulsion speeds and densities. Therefore, we refer to  it as MIPS-like state.

   At low self-propulsion speeds, typically  $Pe \lesssim 10$, the system falls into a collapsed state for almost all packing fractions as the attractive phoretic forces dominate the active force. The threshold $Pe_c$, where the active particles are able to escape, can be estimated by balancing the dimensionless self-propulsion speed and phoretic drift velocity at the contact interparticle distance $r_c=2^{1/6} \sigma$, giving rise to $Pe_c=\zeta_{\text{tr}}/r_c^2 \approx  12$. This estimate is compatible with the results of our numerically obtained state diagram for a large range of packing fractions.
  
  At moderate activities $10\le Pe \le 18$, the system undergoes a transition from an active gas to a MIPS-like state for a large range of packing fractions. Here, MIPS-like state occurs at notably lower densities and self-propulsion speeds than for purely repulsive ABPs, highlighting the role of long-range phoretic interactions in inducing the phase separation.  Especially  at $Pe=10, \, 12$, phase separation occurs at packing fractions as low as  $\Phi=0.02$, see the inset of the phase diagram in Fig.~\ref{phase-diagram_2}.
At higher activities $Pe > 18 $, the system first undergoes a transition from an active gas to a dynamical clustering state at low packing fractions and then to a MIPS-like state at moderate packing fractions, which are still lower than 
the $\Phi$ where purely repulsive active Brownian particles undergo phase separation.

Our results show that attractive long-range chemo-phoretic interactions shift the onset of motility-induced phase separation to remarkably lower packing fractions. Having discussed the overall features of the phase diagram, next we focus on quantifying the structural and dynamical signatures of each  steady-state in the subsequent subsection.
\begin{figure}
\centerline{\includegraphics[width=0.45\textwidth]{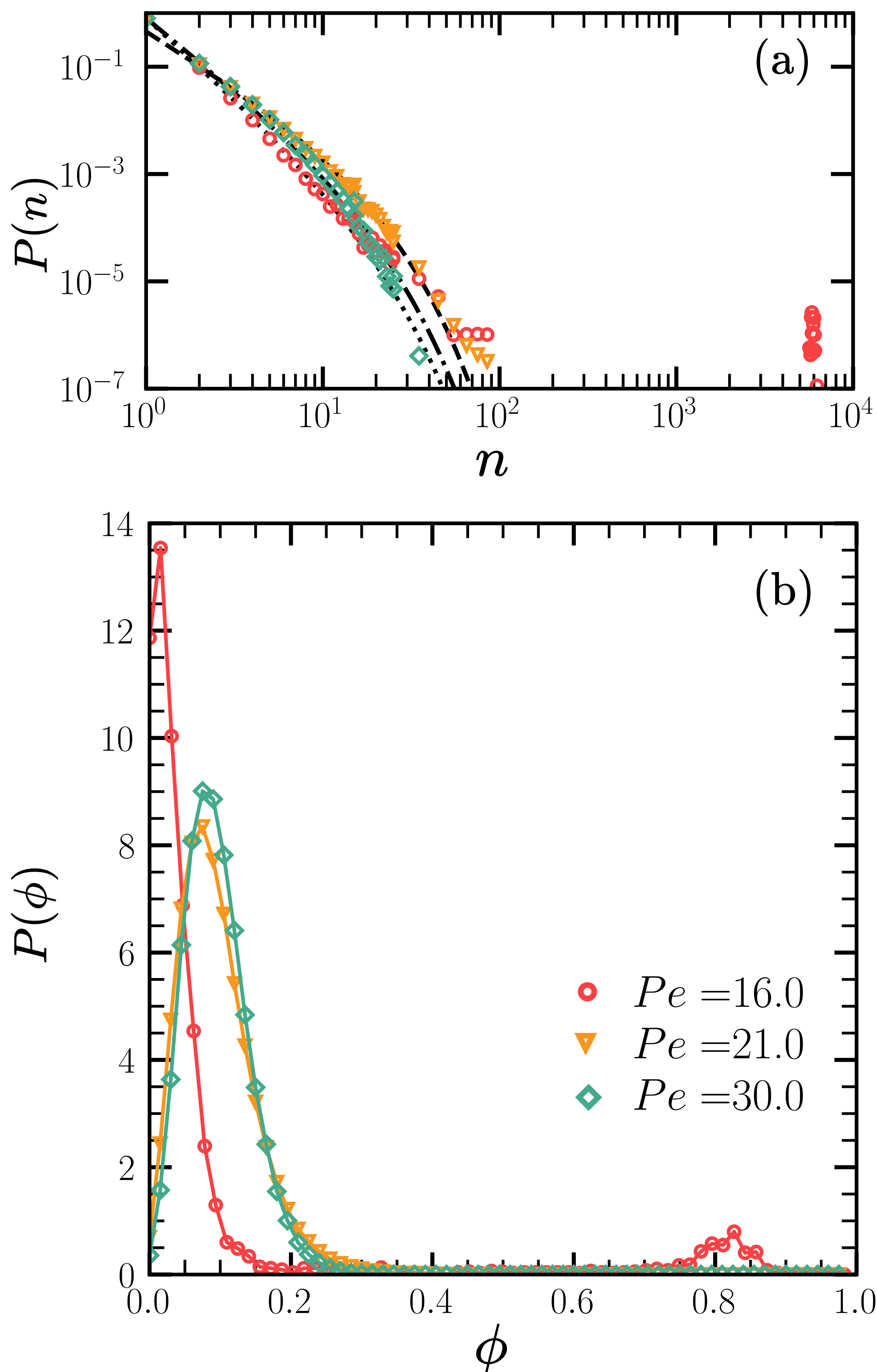}}
		\caption{Probability distribution function of (a) cluster-size  $P(n)$ and (b) local packing fraction $P(\Phi)$ for active colloids of overall packing fraction $\Phi=0.1$ at $Pe=16$ corresponding to a phase-separated state with mean cluster size $ N_{\text{c}}^{\text{avg}}\approx4000$, $Pe=21$ forming dynamical clusters  with $ N_{\text{c}}^{\text{avg}}\approx4$ and $Pe=30$ in the gas state with $ N_{\text{c}}^{\text{avg}}\approx2.8$. The dashed lines in panel (a) correspond to the fits of $P(n)$ with the function $a_0\,n^{-\beta} \,e^{(-n/n_0)}$ with $a_0=0.83$, $\beta = 2.9 $ and $n_0=10$ for the active gas with $Pe=30$, $a_0=0.5$ $\beta = 2.01 $ and $n_0=10$ for the dynamic clusters with $Pe=21$ and $a_0=0.89$, $\beta = 2.6 $ and $n_0=10$ for the active gas part of phase-separated state with $Pe=16.0$. 
		%\daniel{I have added the fitting. Federico's had a mistake in the formula, the minus sign was missing in the exponential. Here are the values $a$=[0.83, 0.5, 0.89], $\beta$=[2.9, 2.01, 2.6] and $n_0$=[10,10,10]. FYI: fitting the data is tricky so what I did is to fix some of the values for Pe=30.0 and once it was fit, free the other ones until a good match. Then for the other Pe' I used the previous values letting $\beta$ and $n_0$ to move}
		} 
	\label{Pn}
\end{figure}
\subsection{Signatures of dynamical states }

In addition to the visual distinction among various dynamical states, we use a set of quantitative measures to distinguish them unambiguously. The quantities include the cluster size distribution $P(n)$, the time-averaged mean cluster size $ N_{\text{c}}^{\text{avg}}$, the fraction of particles in the largest cluster $N_{\text{c}}^{\text{max}}/N$,   the distribution of local packing fraction in the box $P(\phi)$ and the hexatic order parameter $q_6$.  Below, we have summarized the distinctive features of each steady-state.

{\bf Active gas:}
  Consistent with prior studies,~\cite{stark1,stark2} we define an active gas  as a state  where the mean cluster size is smaller than three, $ N_{\text{c}}^{\text{avg}} <3$. The probability distribution of clusters in an active gas can be well described by a power-law exponential curve of the form $a_0 n^{-\beta} \exp(-n/n_0)$ where  $\beta > 2.5$ typically; see Fig.~\ref{Pn}(a) for the example of $\Phi=0.1$ and $Pe=30$. The probability distribution function of the local density also shows a single peak around $\phi \sim 0.1$ as can be seen from Fig.~\ref{Pn}(b).\\

{\bf Collapsed state:}
In this state, the system collapses into a single large cluster similar to the chemotactic collapse that occurs in bacterial systems.~\cite{stark1}
This state corresponds to the case where $ N_{\text{c}}^{\text{max}}/N \to 1$. The haxactic order parameter value in this state is typically  $ q_6 > 0.8 $.

{\bf Dynamical clustering:}
We define the dynamical clustering state as a gas of motile clusters with the  minimal mean cluster of three, {\it i.e.} $ N_{\text{c}}^{\text{avg}} \ge 3$, similar to the definition in prior studies.~\cite{stark1,stark2}  In this state, motile clusters form that strongly fluctuate in shape and size. They ultimately disappear while new ones are formed. The distribution of clusters in this case can also be well described by $P(n)=a n^{-\beta}\exp(-n/n_0)$ with $\beta \sim 2$. See Fig.~\ref{Pn} for an example of cluster size distribution of dynamical clusters with $\Phi=0.1$ and $Pe=21$.
In reference~\cite{stark1}, two different regimes (I and II) of dynamic clustering  were defined.  In the regime I,  $3 \le  N_{\text{c}}^{\text{avg}} \le 6.5$ and
$P(n)$ was found to be described by a single power-law exponential function.
In the regime II, $ N_{\text{c}}^{\text{avg}} > 6.5$ and the $P(n)$ was found to be described by a sum of two power-law exponential functions.  Here, we do not make a distinction between these two dynamic clustering regimes. We simply define a dynamic clustering state as a state for which $  N_{\text{c}}^{\text{avg}}\ge 3$ and the distribution  of the local packing fraction  $P(\phi)$ displays a single peak. In a dynamic clustering state, typically $ N_{\text{c}}^{\text{max}}/N < 0.5$

{\bf Phase-separated state:}
In this state, which we also refer to as MIPS-like, the system phase separates into a large fluctuating cluster which coexists with a dilute fluid which can  be an active gas or a dynamical clustering state. In a phase-separated state,   the distribution of cluster size $P(n)$ in addition to a broad distribution of small clusters, also displays a single peak at large cluster sizes; see Fig.~\ref{Pn}(a) which shows $P(n)$ for the case $\Phi=0.1$ and $Pe=16$. In this  case, the distribution of small clusters is very similar to that of an active gas. Therefore, a single giant cluster coexists with an  active gas. To quantify our visual observation of phase separation, we look into the probability density of the local packing fraction $P(\phi)$ as shown in Fig.~\ref{Pn} (b)  for $\Phi=0.1$ and $Pe=16$ where we observe a double peak distribution. The first peak lies at the low packing fraction $\phi_1\sim 0.02 <0.1$ and the second peak is at a high packing fraction around $\phi_2 \sim 0.85  \gg 0.1$, confirming the coexistence of two fluids, one dilute active gas and a dense large cluster.

\begin{figure}[h!]
	 \centerline{\includegraphics[width=0.95\linewidth]{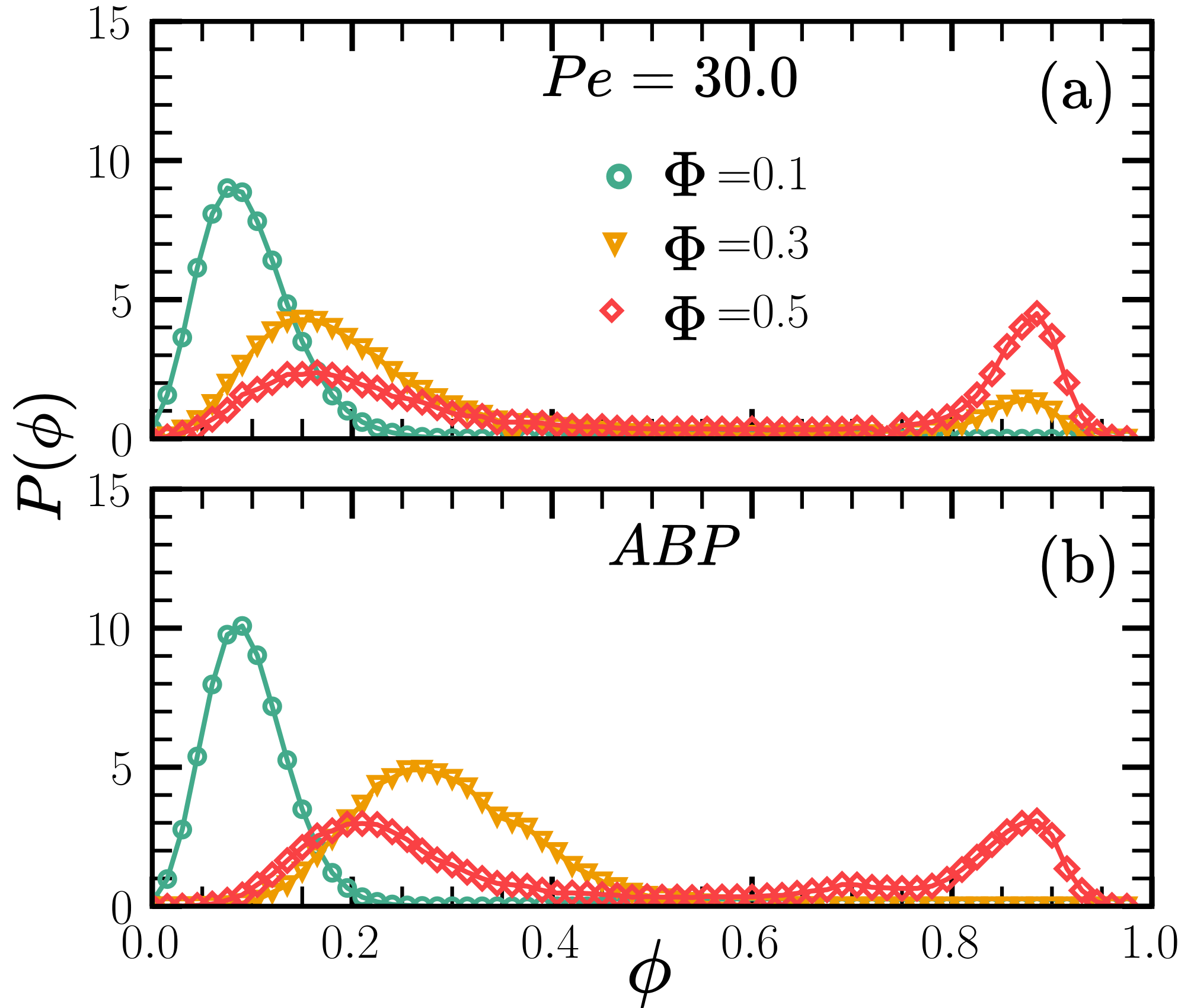}}
	\caption{Probability distribution of local density $P(\phi)$ at $Pe=30$ for (a) self-phoretic colloids at $\Phi=0.1$ (active gas), $\Phi=0.2$ (dynamical clustering) $\Phi=0.3$ and 0.5  (phase-separated states), respectively. (b) active Brownian particles at $\Phi=0.1$ and 0.3 (active gas) and $\Phi=0.5$ and 0.7  (motility-induced phase-separated states).
}
	\label{dens_prob}
\end{figure}

The two-peak distribution function for $P(\phi)$ is a generic feature for all phase-separated states independent of the $Pe$ and $\Phi$ values, see Fig.~\ref{dens_prob}(a) for a few examples of the case $Pe=30$. This is very similar to what is observed for MIPS  in purely repulsive ABP systems. For comparison, we have shown the $P(\phi)$ of ABP system for for $Pe=30$ at $\Phi=0.1$,0.3 and 0.5 in Fig.~\ref{dens_prob}(b). For $\Phi=0.5$ where phase separation occurs, we observe a two-peak distribution function.
For self-phoretic active colloids at $Pe=30$, the probability distribution function switches from a single-peak to double-peak function for $\Phi>0.2$ where we see a low-density peak around $\phi \sim 0.15$ and a high-density peak  around  $\phi_2> 0.8$, where the value of $\phi_2$ approaches the close packing density $\phi \approx 0.9$. 
%For comparison, we have also included the  $P(\phi)$ of ABP system for the same  $Pe=30$ and several  packing fractions.
For ABPs, the transition of  $p(\phi)$ from single-peak to double-peak occurs at higher densities  $\Phi > 0.3$ compatible with prior studies.~\cite{hagan}

\begin{figure*}
	%\centerline{\includegraphics[width=0.5\textwidth]{figures/hexatic.png}}
    \includegraphics[width=1.0\linewidth]{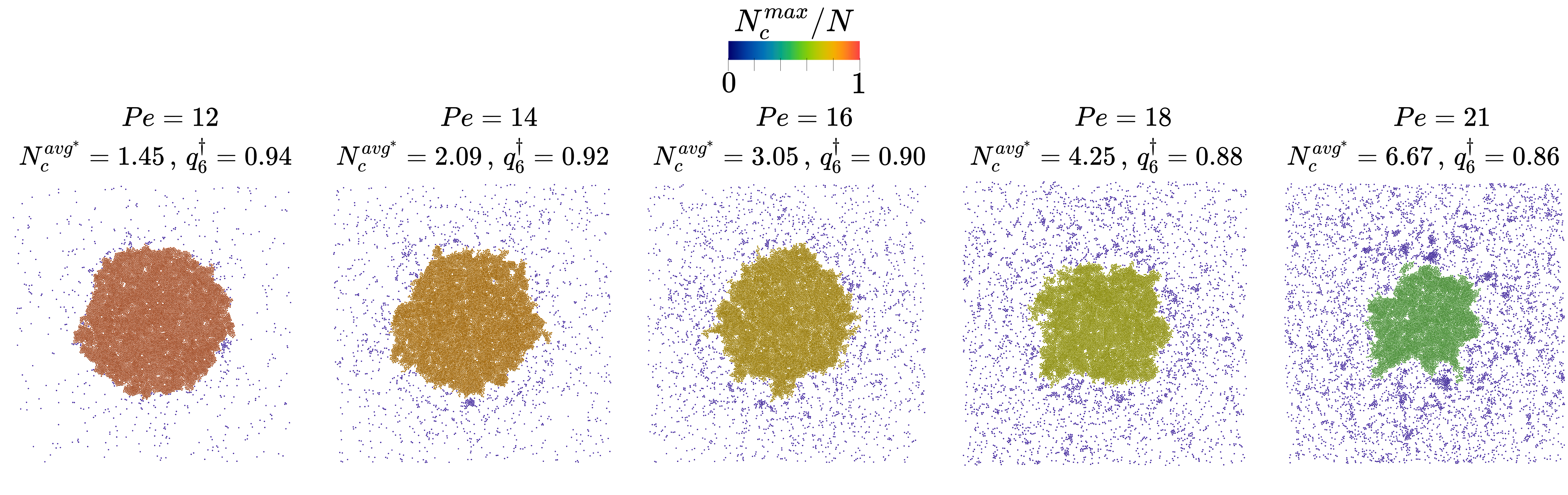}
	\caption{Chemotactic phase-separated states found at  packing fraction $\Phi=0.2$. Here, $ N_{\text{c}}^ {\text{avg}^*} $ is the average cluster size outside the biggest cluster in the system, and $q_6^{\dagger}$ is the hexatic order parameter inside the biggest cluster in the system. Color codes represents the order parameter $N _{\text{c}} ^{\text{max}}/N$.}
	\label{MIPS-PHASES}
\end{figure*}

The curves of $P(\phi)$ in Fig. \ref{dens_prob} show that in the case of nonequilibrium phase separation, unlike the equilibrium situation, the system does not always  separate to the same dilute and dense phases. In particular, the locations of the peaks do not seem to coincide in Fig. \ref{dens_prob}(a). Indeed, further investigating the phase-separated states at a fixed overall packing fraction $\Phi=0.2$, while varying $Pe$, see Fig.~\ref{MIPS-PHASES}, we can distinguish distinct types of phase-separated states. The dilute phase can be either an active gas or a dynamical cluster as can be deduced by the \emph{mean number of clusters in the dilute phase} (outside of the biggest cluster) denoted by $ N_{\text{c}}^{\text{avg}^*}$. Interestingly, for $Pe=16$ and $18$, the dilute phase is a dynamical cluster whereas single phase dilute systems at the same $Pe$ but lower $\Phi$ is an active gas.
The dense phase can be either a disordered liquid or ordered (hexatic liquid or active solid). To quantify the degree of order within the dense phase, we compute the hexatic order parameter inside the largest cluster in the system, which we denote it by $q_6^{\dagger}$. Upon the increase of $Pe$, the dense phase becomes more disordered as evidenced by the decreasing value of hexatic order parameter within it  $q_6^{\dagger}$, whereas the mean size of clusters in the dilute phase $ N_{\text{c}}^{\text{avg}^*}$ increases.
%\daniel{This need to be clarified, what is $\dagger$ and $\star$}

%In Fig.~\ref{MIPS-PHASES}

%In ABPs MIPS creates a phase separation between a dilute and dense phase which can be liquid, hexactic or solid as the activity is increased.

\section{Characteristics of dynamical phase transitions }
Having provided an overview of the state diagram and salient features of each dynamical state, we focus on understanding the evolution of the relevant order parameters as a function of $\Phi$ for a wide range of self-propulsion speeds $ 5 \le Pe \le 30$. We can recognize three distinct regions in the state diagram:
\begin{enumerate}
    \item At low self-propulsion speeds $ 5 <Pe <10$, the system  goes directly from an active gas to a collapsed state. %\sara{In this case the morphology of collapsed state changes upon increase of $\Phi$??.}
\item For intermediate self-propulsion speeds 
$10 \le Pe < 19$, the  system undergoes a dynamical transition from an active gas to a phase-separated state at intermediate packing fractions and finally to a collapsed state at sufficiently high packing fractions $\Phi \ge 0.50$.

\item For $Pe \ge 19$, we observe a dynamical transition first from the active gas to dynamical clusters, then to a phase-separated state up to highest investigated packing fractions $\Phi=0.7$. 
\end{enumerate}

To investigate the nature of dynamical transitions from active gas and dynamical clusters to the phase-separated state, we probe various structural and dynamical measures as a function of $\Phi$ at different self-propulsion speeds in the range $ 10 \le Pe \le 30$.

\subsection{Mean and maximum cluster size}
\begin{figure}
\centering
\includegraphics[width=0.95\linewidth]{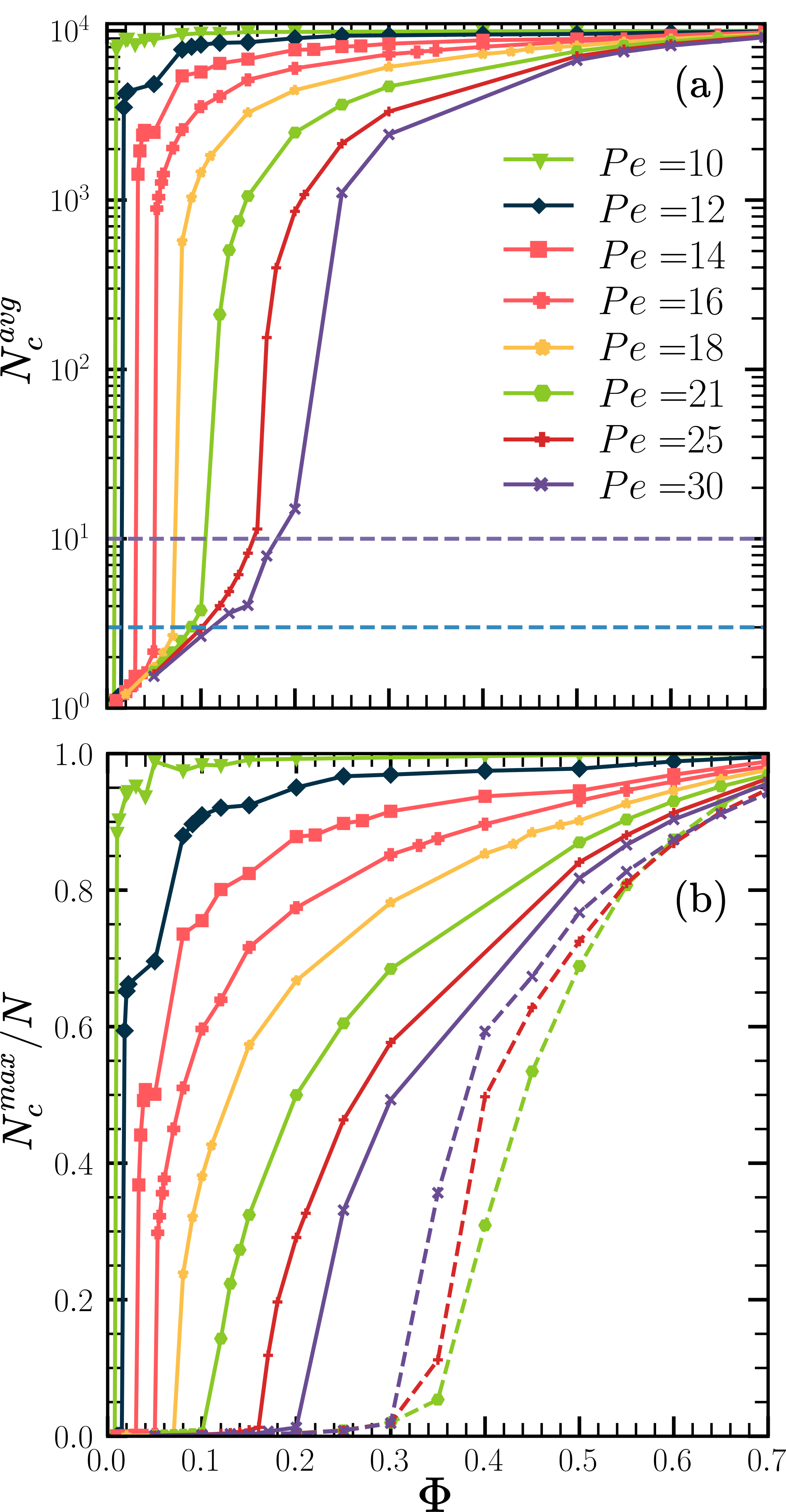}
\caption{ (a) Time-averaged mean cluster size $ N_{\text{c}}^{\text{avg}}$ and (b) Ratio of time-averaged largest cluster size $N _{\text{c}} ^{\text{max}}$ relative to total number of particles  $N=10^{4}$ of self-phoretic colloids ($\zeta_{\text{tr}}=15.4$ and $\zeta_{\text{rot}}=-0.38$) as a function of the packing fraction $\Phi$ for various P\'{e}clet numbers $10 \le Pe \le 30$ as given in the legend. The dashed lines in panel (a) correspond  to $ N_{\text{c}}^{\text{avg}}=3$ and 10 respectively. The dashed lines in panel (b) show  $N _{\text{c}} ^{\text{max}}/N$ for ABP particles with periodic boundary conditions.}
\label{Nc_Nmax} 
\end{figure}

First, we look at the mean size of clusters $ N_{\text{c}}^{\text{avg}}$ and the ratio of the mean largest cluster size to the total number of particles $N _{\text{c}}^{\text{max}}/N$  as as shown in Fig.~\ref{Nc_Nmax} (a) and (b), respectively. %At the lowest self-propulsion speed $Pe=10$ corresponding to the  regime  1, both  $ N_{\text{c}}^{\text{avg}}$ and $N _{\text{c}} ^{\text{max}}/N$ display a sharp transition at a very low packing fraction $\Phi=0.01$, reminiscent of first-order thermodynamic transition.
For P\'{e}clet numbers in the range $10 \le Pe \le 18$, we observe a sharp transition of $ N_{\text{c}}^{\text{avg}}$ from small values $ N_{\text{c}}^{\text{avg}} <3 $ to large values $ N_{\text{c}}^{\text{avg}} > 1000 $ at a low packing fraction $\Phi_L < 0.1$ which its value increases with the self-propulsion speed $Pe$. Especially, at the lowest self-propulsion speed $Pe=10$  the sharp transition occurs at a very low packing fraction $\Phi=0.01$. Interestingly, the order parameter $N _{\text{c}} ^{\text{max}}/N$
 also shows an abrupt jump at the same  low $\Phi_L< 0.1$  followed by a second jump at   higher densities $ \Phi_H > 0.1$ for $Pe \le 16$, reminiscent of first-order thermodynamic transitions. The first jump corresponds to a transition from an active gas to a phase-separated state, whereas the second jump demarcates a transition to a collapsed state. For $Pe >20$, $ N_{\text{c}}^{\text{avg}}$ changes more continuously.  Nevertheless, we observe a steep increase in $ N_{\text{c}}^{\text{avg}}$  around $\Phi=0.1$ where the system goes from an active gas   to a dynamical clustering state for which $ N_{\text{c}}^{\text{avg}} > 3 $. We note that in this regime  $N _{\text{c}} ^{\text{max}}/N$  remains zero for $\Phi<0.2$ and afterwards  when the system goes into a phase-separated state, it evolves continuously akin to a second-order thermodynamic phase transition. Our findings suggest that $N _{\text{c}} ^{\text{max}}/N$ is a good order parameter characterizing the transition from either an active gas or dynamical clustering state to a phase-separated state.
%I think we need more data for Pe=21 for phi>0.1

\subsection{Hexatic order parameter}
\begin{figure}
	%\centerline{\includegraphics[width=0.5\textwidth]{figures/hexatic.png}}
    \includegraphics[width=0.95\linewidth]{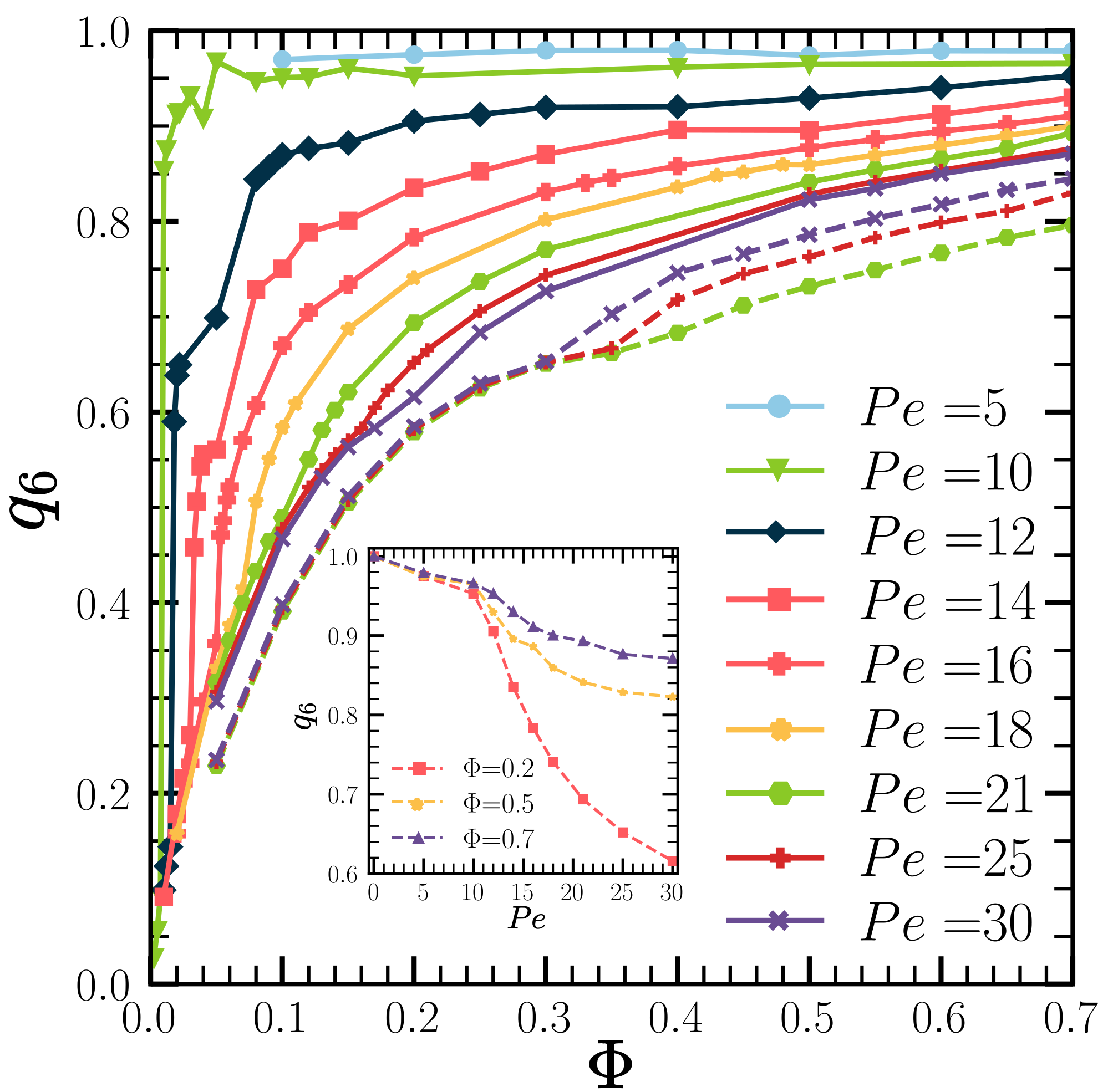}
	\caption{Time-averaged $6$-fold-bond orientational order parameter $q_{6}$ as function of the packing fraction $\Phi$ for $Pe=21, 25$ and $30$ and $\zeta_{\text{tr}}=15.4$ and $\zeta_{\text{rot}}=-0.38$ for chemotactic particles ($\zeta_{\text{tr}}=15.4$ and $\zeta_{\text{rot}}=-0.38$) and for ABP particles in periodic system. Inset shows the $q_6$  for different values of $Pe$ at $\Phi=0.2$ and $\Phi=0.5$. }
	\label{hexatic}
\end{figure}

We have also investigated the   global six-fold bond orientational order parameter $q_6$,  also known as hexatic order parameter, as a function of packing fraction at different  P\'eclet numbers.  Fig.~\ref{hexatic} shows the time-averaged values of the $q_6$ order parameter as a function of the packing fraction $\Phi$ at various P\'eclet numbers for self-phoretic colloids with $\zeta_{\text{tr}}=15.4$ and $\zeta_{\text{rot}}=-0.38$ (continuous lines).

The general trend that we observe is that $q_{6}$ increases with $\Phi$ at each self-propulsion speed. However, the higher $Pe$ the value of $q_6$ at identical packing fractions is lower. We note that for low self-propulsion speeds $Pe=5$ and 10 where the system is in a collapsed state already at packing fractions as low as $\Phi=0.1$, $q_6$ is very close to unity,  a clear evidence of an overall hexatic order.  At this stage, we cannot tell definitely if the system is an active solid or a hexatic liquid as clarifying this requires the calculation of spatial density correlations for very large systems beyond the system size investigated here.~\cite{Bernard2011,rene3,MIPS-6} However, visual inspections suggest that we have an active solid with long-range positional order.

For  $12 \le Pe \le 16$, where the system undergoes a transition from an active gas to a phase-separated state and finally to a collapsed state, we observe a steep increase of $q_6$ upon phase separation of the system into  dense and dilute fluids and a second remarkable increase of $q_6$ when the system enters the collapsed state. The observed trends reinforce the idea that the transitions from an active gas to phase-separated state and then to a collapsed state are first-order dynamical transitions. For larger self-propulsion speeds, $Pe>18$, where the system undergoes a transition from the active gas to dynamical clustering state, then to the phase-separated state, $q_6$ changes continuously with $\Phi$ again consistent with the trends observed for $N _{\text{c}} ^{\text{max}}/N$.

For comparison, we have also included the hexatic order parameter of the ABP system with periodic boundary conditions at   $Pe=21,25$ and 30 shown by dashed lines. We note that ABPs overall show a weaker degree of hexatic order than chemotactic colloids at identical values of $Pe$ and $\Phi$.  The value of hexatic order parameter in all collapsed states is remarkably high, $q_6 >0.85$, comparable to $q_6$ values obtained in active solids of the ABP system.~\cite{MIPS-6} This strongly supports our inference that the collapsed state is an active solid.  Moreover, if we plot $q_6$ as a function of $Pe$ at a fixed $\Phi$ as shown in the inset of Fig.~\ref{hexatic} for $\Phi=0.2$, 0.5 and 0.7, we observe a notable reduction of $q_6$ at the $Pe$ values where the system enters a phase-separated state. 
 %A similar behavior for the ABP particles.

 \subsection{Giant fluctuations}

One important signature of the nonequilibrium nature of active systems is captured by the giant number fluctuations. In an equilibrium system, the standard deviation $\Delta N$ of the mean number of particles $\bar{N}$ in a subvolume scales as $\Delta N \sim \bar{N}^{1/2}$ for $\bar{N} \rightarrow \infty$. However, in an active system this is not true anymore and $\Delta N$ can scale differently as $\Delta N \sim \bar{N}^{\alpha}$ with $ 1/2<\alpha \le 1$.  Although giant number fluctuations were originally predicted in active systems with nematic interactions, the authors in reference~[\onlinecite{Fily2012}] reported the existence of giant number fluctuations in a system of self-propelled disks, without nematic alignment interactions, undergoing phase separation. It turns out that giant number fluctuations are a universal feature of active systems.~\cite{Dey_2012} To see the effect of chemotactic interactions on density fluctuations, we computed the exponent $\alpha$ of giant number fluctuations in subsystems with linear dimensions $L/n$, where $n=4,8,16,32,64$.  In each case, we  evaluated the mean number of particles $\bar{N}$  and its standard deviation $\Delta N$ within each subsystem. 

\begin{figure}
	\centerline{\includegraphics[width=0.95\linewidth]{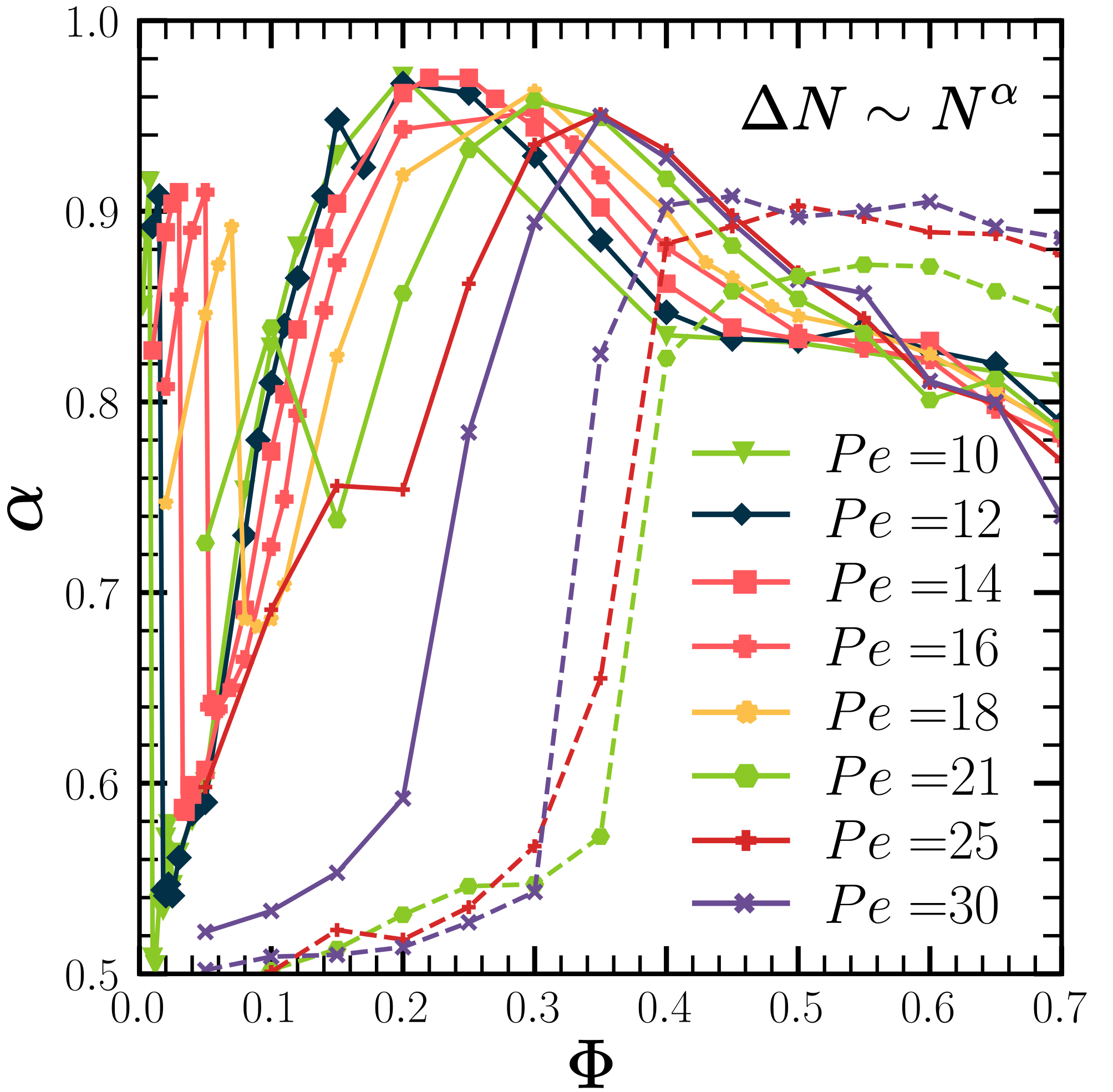}}
	\caption{Plot of the exponent $\alpha$ of the relation $\Delta N \sim \bar{N}^{\alpha}$ as function of the packing fraction $\Phi$ for $Pe=16, 21, 25$ and $30$ with $\zeta_{\text{tr}}=15.4$ and $\zeta_{\text{rot}}=-0.38$ (continuous lines) and for ABP in periodic system (dashed lines).}
	\label{giant}
\end{figure}

Fig.~\ref{giant} shows the exponent $\alpha$  as function of the packing fraction $\Phi$ for various P\'{e}clet numbers in the range $ 10 \le Pe \le 30$   for chemotactic colloids with $\zeta_{\text{tr}}=15.4$ and $\zeta_{\text{rot}}=-0.38$ (continuous lines) and for ABP system in periodic box (dashed lines). Let us first focus on  the  region of the phase diagram,  $10\le Pe <18$, where the system undergoes two dynamical transitions; first from an active gas to a phase-separated state and then to a collapsed state.
For the active gas state  $\alpha$ increases systematically with $\Phi$ from $1/2$ to very large values $\alpha \approx 0.9$ until the system undergoes a phase separation where $\alpha$ drops to values below 0.6. Within the phase-separated regime, however, $\alpha$ keeps on increasing with $\Phi$ to very large values $\alpha \approx 0.95$   until about $\Phi \sim 0.2$ where $\alpha \approx 0.95$. Afterwards, we observe a decline of $\alpha$ until $\Phi \approx 0.4-0.5$, where the system enters a collapsed state. In the collapse state, it remains nearly constant  $\alpha \approx 0.4$.

 For higher P\'eclet numbers $Pe > 18$, we observe an increase of $\alpha$ with $\Phi$ as the system goes from active gas to dynamical clustering state and then to a phase-separated state up to $\Phi \sim 0.3$.  Beyond this point, $\alpha$ decreases  with $\Phi$ although it remains significantly larger than $0.5$ for all packing fractions.  The decrease of $\alpha$ beyond $\Phi \sim  0.3$ can be understood in view of increasing size of the largest cluster for large $\Phi$. For $\Phi \sim  0.3$ the system is already in a MIPS-like state in which a central giant cluster coexists with smaller dynamical clusters and particles continuously join  and leave the big cluster. The higher the density, the bigger the central cluster and  the smaller the number fluctuations associated with particles joining and leaving the big clusters. This translates into a reduced exponent $\alpha$.
 
 In the case of purely repulsive ABPs,  $\alpha \sim 0.5$ for $\Phi < 0.2$  close to its value for the equilibrium case. Upon further increase of packing fraction, we observe a notable increase of $\alpha$ in the region $ 0.2<\Phi \lesssim 0.4$ concomitant with the occurrence of the motility-induced phase separation. For $\Phi >0.4$, the exponent saturates to a value $ \alpha \sim 0.9$  in agreement with prior results in the literature.~\cite{Fily2012,marchetti3}
 Comparing the  number fluctuation exponents of chemotactic and active Brownian colloids at high densities  reveals that the exponent $\alpha$ of ABPs with identical  $\Phi$ and $Pe$ are larger. This indicates that attractive chemotactic interactions reduce the density fluctuations by forming larger clusters as is also visible in larger values of  $N _{\text{c}} ^{\text{max}}/N$ for the chemotactic system, see Fig.~\ref{Nc_Nmax} (b). The decrease of $\alpha$ with $N _{\text{c}} ^{\text{max}}/N$ in the phase-separated region is also visible when we consider a fixed $\Phi$ and increase $Pe$. Let us consider $\Phi=0.2$ for which we observe a decrease of $\alpha$ with increasing $Pe$. If we compare this trend with the evolution of phase-separated state with $Pe$ presented in Fig.~\ref{MIPS-PHASES}, we note that as we increase $Pe$, the dilute phase evolves from an active gas to a dynamical clustering state whereas the mean size of the big cluster decreases. In other words, the contrast between the dilute and dense phases decreases with $Pe$, leading to a decrease of density fluctuations and therefore the exponent $\alpha$.

\section{Concluding remarks}

We have explored the dynamical  steady-states of a monolayer of chemotactic self-phoretic active colloids employing Brownian dynamics simulations, in which the particles translational and rotational degrees of freedom are coupled to the chemical field.  Assuming the chemical field diffuses much faster than the colloids, we adopted a stationary profile of the chemical field.~\cite{stark1,stark2} We studied the state diagram of the system for the case that chemical-mediated interactions induce effectively attractive forces and repulsive torques. We have investigated the features of dynamical states  emerging from the interplay between steric-chemical interactions and self-propulsion for  a wide range of packing fractions and  self-propulsion speeds encoded by the P\'{e}clet number ($Pe$).

We find four distinct dynamical steady-states: (\emph{i}) the collapsed state in which all particles join a giant cluster occurring at relatively low self-propulsion speeds $Pe \lesssim 10$ or intermediate $Pe$ for dense systems $\Phi > 0.6$ (\emph{ii}) an active gas appearing at  $Pe \ge 10 $ for relatively low packing fractions $\Phi \lesssim 0.1$, (\emph{iii}) a 
 dynamical clustering state for $0.05 \lesssim \Phi < 0.2$ and  $Pe >18$, and (\emph{iv}) a phase-separated state where a big cluster coexists with a dilute gas  reminiscent of the Motility-Induced Phase Separation (MIPS) reported in  purely repulsive active system.   
 The  phase-separated state is well revealed by  the probability distribution functions of of clusters and local  packing fraction.  When the system phase separates,  both distribution functions display a secondary peak. 
 
 To our knowledge, our work is the first report of a MIPS-like state in chemotactic colloids. What is remarkable is that the long-range phoretic interactions shift the onset of MIPS to much lower packing fractions. The MIPS-like state can occur at packing fractions as low as $\Phi \approx 0.01 $ for intermediate self-propulsion speeds. There have been reports of the formation of low density three-dimensional living clusters~\cite{saric} in self-propelled particles interacting via an attractive Lennard-Jones potential, but they resemble the dynamical clustering state which is also reported for low density, $\Phi \sim 0.05$, chemotactic colloids.~\cite{stark1,liebchen3}  In reference~\cite{hagan2} a   2D active colloid system at $\Phi=0.4$ with full Lennard-Jones interaction was investigated. It was shown that the attractive part of the Lennard-Jones forces can enhance the cluster formation at low $Pe$ and induce a re-entrant phase separation at high $Pe$. However, the  low-$Pe$ phase-separated states look like a system-spanning colloidal gel, whereas the kinetics of phase separation in MIPS-like states of chemotactic colloids presented here are governed by nucleation, growth, and coarsening for all the investigated range of $Pe$ and $\Phi$.

 Our study also shows that the transition from an active gas or dynamical clustering state to the MIPS-like state upon increase of packing fraction can be well captured by the fraction of particles in  the largest cluster $ N_{\text{c}}^{\text{max}}/N$.   At intermediate $Pe$ where the system undergoes a transition from an active gas to MIPS-like state, $ N_{\text{c}}^{\text{max}}/N$ displays an abrupt jump   reminiscent of a first-order  thermodynamic phase transition.  
 At higher $Pe$ where the system enters a phase-separated state from a dynamical clustering state, $ N_{\text{c}}^{\text{max}}/N$ changes continuously with $\Phi$ akin to a second-order thermodynamic phase transition. Overall, chemotactic colloids form smaller clusters when the self-propulsion speed is increased. However, compared to the case of purely active Brownian particles, chemical field-induced attractions shift the onset of cluster formation and phase separation to lower densities and induce a larger hexatic order parameter at identical $Pe$ and $\Phi$.
 
 %At low packing fractions, however, the dynamical clustering region seems to disappear. As a well-established signature of the active nature of the studied system, we also checked that the systems exhibit giant number fluctuations and compared that with the case of purely ABP. Differently from the ABP case, chemotactic particles show higher values for the exponent and a more pronounced decreasing behavior for very high densities.\\
 
To conclude, our results for a stationary chemical field highlight the role of long-range chemical field-mediated interactions on  inducing phase separation of self-phoretic active colloids at  relatively low  self-propulsion speeds and remarkably low packing fractions.
In a study where the full time-dependent solution of the chemical field coupled to the orientational degrees of phoretic colloids was considered,~\cite{liebchen3} interesting wave patterns resulting from delay effects emerged.
It remains open what dynamical patterns unfold when both translational and rotational degrees of freedom are coupled to non-stationary chemical field gradients and denser systems.  In the future, we plan to extend this work to consider an explicit solution of the dynamical equation of the chemical field, where similar to references~\cite{liebchen3,nejad} screening of the chemical field is directly implemented. Eventually, it will be of interest to  introduce chirality into the equations of motion and investigate the interplay between chiral and chemotactic interactions.

\section{Acknowledgments}

The authors warmly thank Holger Stark for  helpful discussions. F.F. acknowledges the funding from the Delta Institute of Theoretical Physics and  Lorenzo Caprini for enlightening discussion and suggestions. D.A.M.F would like to thank Demian Levis for his valuable insights into motility-induced phase separation. This work was part of the D-ITP consortium, a program of the Netherlands Organization for Scientific Research (NWO) that is funded by the Dutch Ministry
of Education, Culture and Science (OCW).  
The computations were carried out on the Dutch national e-infrastructure with the support of SURF Cooperative. F.F and D.A.M.F contributed equally to this work. 

\bibliography{manuscript}

\end{document}